# A New Interpretation of Two-Photon Entangled Experiments via Quantum Mirrors

D. B. Ion, P. Constantin and M.L.D. Ion

National Institute for Physics and Nuclear Engineering, Horia Hulubei IFIN-HH,
P.O.Box MG-6, Romania

**Abstract**: In this paper the coherence and crossing symmetry of the electromagnetic interaction of photons involved in the spontaneous parametric down conversion (SPDC) phenomena in nonlinear crystals, are investigated. On this basis a new interpretation of the recent results on the two photon entangled experiment is obtained. A new field of applications is suggested.

## 1. Introduction

The spontaneous parametric down conversion (SPDC) is a nonlinear optical process [1] in which a laser pump beam incident on a nonlinear crystal leads to the emission of a correlated pair of photons. In this process energy and momentum of photons are conserved. The essential quantum nature of the resulting two-photon correlated states has been confirmed by many interesting experiments [2]. Recently [3] the SPDC process allowed to demonstrate two-photon "ghost" imaging and interference-diffraction patterns as well as other new phenomena from the geometric and physical optics. In this paper the coherence and crossing symmetry of the electromagnetic interaction of photons involved in the SPDC processes are investigated. So, a new and complete interpretation of the "ghost " two photon image and diffraction patterns, reported in Ref. [3], is obtained in terms of the crossing symmetric processes in SPDC crystals.

## 2. Two-photon ghost imaging and interference diffraction experiments

The results of two experiments with SPDC light beams reported in Ref. [3] are as follows. The SPDC light beam, consisting of pairs of orthogonally polarized photons, is splitted into two diverging beams (called signal (s) and idler (i) ) by a polarization beam splitter (BS) so that coincidence detections may be performed between two distant photon-counting detectors.

*2.1 Two-photon optical imaging experiment*

In the two-photon ghost image experiment [3,4] an argon ion laser is used to pump a nonlinear BBO crystal ( $\beta - BaB_2O_4$ ) to produce pairs of orthogonally polarized photons (see Fig. 1 in Ref. [3] for detailed experimental setup). After the separation of the signal and idler beams, an aperture (mask) placed in front of one of the detectors ( $D_1$ ) is illuminated by the signal beam through a convex lens. The surprising result consists from the fact that an image of this aperture is observed in coincidence counting rate by scanning the other detector ( $D_2$ ) in the transverse plane of the idler beam, even though both detectors' single counting rates remain constants. For understanding the physics involved in this experiment in Fig. 1 we illustrated a simplified scheme (by removing the signal detector $D_1$ as well as the collection lens) of the 'unfolded" version of the two-photon imaging setup (see Fig. 3 in Ref. [3]). The remarkable feature in this experiment is the validity of the Gaussian thin-lens equation

$$\frac{1}{S} + \frac{1}{S'} = \frac{1}{f} \qquad (1)$$

which is satisfied to a surprising accuracy by the focal distance f of lens, the distance S from lens to aperture and the distance S` from lens to "ghost " image in $D_2$ via BBO crystal. In other words the results obtained in this experiment can be summarized as follows:

*(R.1) The image is exactly the same as one would observe on a screen placed at the fiber tip if the detector $D_1$ were replaced by a pointlike light source and the SPDC BBO crystal by a reflecting mirror.*

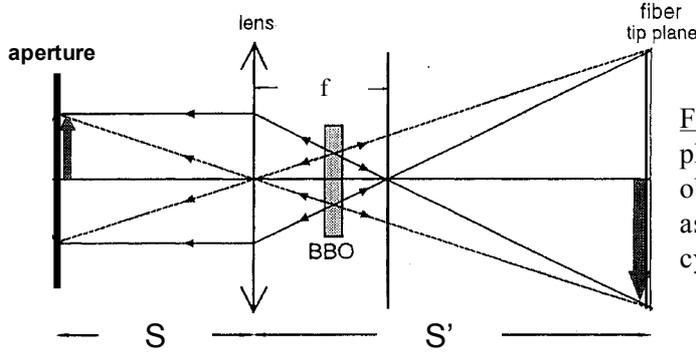

Fig.1. A conceptual *unfolded* version of the two photon *ghost* imaging experiment [3]. As one observe our scheme does not contain the detector as pointlike („ghost") source one prove that BBO cystal acts as a real quantum mirror (see R6).

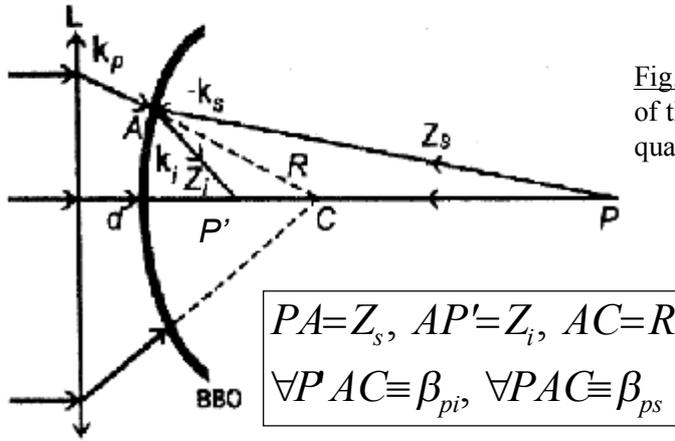

Fig.2. The basic optical configuration for a proof of the radial focal law (14) of the spherical quantum mirror.

*2.2 Two-photon "ghost " diffraction experiment*
The experimental setup (see Fig. 4 in Ref. [3]) is similar to that used in "ghost " image experiment except that rather than an aperture it is a Young's single slit (or double slit) inserted into the path of the signal photons beam. Surprisingly, a diffraction (or an interference pattern) is observed in the coincidences measurements when scanning the detector $D_2$ in the idler beam even the single-detector-counting rates are both observed to be constant when scanning detector $D_1$ in signal beam or $D_2$ in the idler beam. This counting rate in single-slit diffraction pattern ,obtained in Ref. [3], is described by the usual relation

$$\frac{R_c(x_2)}{R_c(0)} = \left[\frac{\sin X}{X}\right]^2 , X \equiv \frac{\pi a x_2}{\lambda z_2} \qquad (2)$$

where a is the slit width, while $x_2$ and $z_2$ are the distances defined in Fig.7 from Shih et al [3]. Therefore, the results obtained in two-photon "ghost "diffraction experiment [3] can be summarized as follows:
(R.2) *The diffraction pattern is the same as that which one would observe on a screen in the plane of $D_2$, if $D_1$ detector is replaced by a pointlike source and the SPDC-BBO crystal by a reflecting mirror as in Fig. 2.*

It was recently shown [5] that the down-converted light has similar coherence area properties as the ones of incoherent light source. Moreover, it was also proved [6] that interference patterns can be detected performing coincidence measurements between a conjugated signal and idler pairs. The main results of these experiments are as follows:

(R.3) *Interference patterns were produced [5] in double-slit experiments with visibility controlled by the distance between source and slits, despite the high degree of directionality (- 1 mrad ) of the down converted light around of a given wavelength.*
(R.4) *The degree of visibility of interference patterns obtained [6] with a double-slit placed at the signal beam can be nonlocally controlled through the idler beam.*
(R.5) *The visibility of interference fringes produced by a signal beam transmitted through a double slit, can also be controlled [7] by aligning an auxiliary laser with the idler beam, with the same wavelength and varying its power (see Figs. 2a,b in Ref.[7]). In this case, the degree of coherence of the source is varied directly by the inducing laser intensity without performing any measurements on the idler beam.*

### 3. A new interpretation of the two-photon entangled experiments

We consider that the key for the understanding of all above results (R.1)-(R.5) is given by the following two distinctive features of the SPDC crystals.

3.1. *The Cherenkov-like coherence condition of the SPDC-photons.*
Indeed, a coherence condition, similar with that from the usual Cherenkov effect [8], can be proved by using energy-momentum conservation relations ( $\eta = c = 1$ ) in the SPDC crystals

$$\omega_p = \omega_s + \omega_i, \quad \vec{k}_p = \vec{k}_s + \vec{k}_i \tag{3}$$

Hence, let $\theta_{ps}$ (or $\theta_{pi}$) be the angle between the incident and signal ( and idler) photons and let $v_{ph(s)}$ and $v_{ph(i)}$ be the corresponding phase velocities and $v_0$ the pump photon group velocity. Then, from Eq.(3) we get the following Cherenkov-like coherence conditions:

$$\cos\theta_{ps} = \frac{k_p^2 + k_s^2 - k_i^2}{2k_p k_s} \leq 1 \text{ or } \cos\theta_{ps} \approx \frac{v_{ph(s)}}{v_0} \leq 1 \tag{4a}$$

or

$$\cos\theta_{pi} = \frac{k_p^2 + k_s^2 - k_i^2}{2k_p k_s} \leq 1 \text{ or } \cos\theta_{pi} \approx \frac{v_{ph(i)}}{v_0} \leq 1 \tag{4b}$$

Therefore, in the signal beam ( or in the idler beam) the photons can have high coherence properties similar with that from the usual Cherenkov effect.

3.2. *Crossing symmetric SPDC processes and quantum mirror*
If the S-matrix crossing symmetry [9] of the electromagnetic interaction in the SPDC crystals is taken into account, then the existence of the direct SPDC process

$$p \rightarrow s + i \tag{5}$$

will imply the existence of the following crossing symmetric processes

$$p + \overline{s} \rightarrow i \tag{6}$$

$$p + \overline{i} \rightarrow s \tag{7}$$

as real processes which can be described by the same transition amplitude.
Here, to each signal photon $s(\vec{k}_s, \mu_s)$ [or $i(\vec{k}_i, \mu_i)$] we associated the corresponding phase conjugated mirror (PCM) photon $\overline{s}(-\vec{k}_s, -\mu_s)$ [or $\overline{i}(-\vec{k}_i, -\mu_i)$], respectively, where by $\mu$ we denoted the photon helicity.
Now, instead of a general proof, we present here a brief discussion of the phase conjugation mechanism by which we can understand the existence of the SPDC crossing symmetric processes (6)- (7). Any nonlinear medium illuminated by a high quality optical beam (as is the pump laser beam) can be considered an optical phase conjugation mirror (PCM) [10], by which the wave vector is reversed while the polarization vector e is complex conjugated

$$PCM : (\vec{k}, \vec{e}) \rightarrow (-\vec{k}, \vec{e}^*) \quad (8)$$

Indeed, as was first pointed out by Yariv [12] and proved experimentally by Avizonis et al. [13], one can use the method\ of three-wave mixing to generate phase-conjugate replicas of any optical beam. This scheme exploits the second order optical nonlinearity in a crystal lacking inversion symmetry. In such crystals, the presence of input field

$$E_p = \frac{1}{2} E_p(\omega_p) \exp(i[\omega_p t - \vec{k}_p \cdot \vec{r}]) + c.c.$$

$$E_s = \frac{1}{2} E_{pw}(\omega_s) \exp(i[\omega_{pw} t - \vec{k}_{pw} \cdot \vec{r}]) + c.c. \quad (9)$$

where the $\vec{E}_p$ and $\vec{E}_{pw}$ the pump (p) and probe waves (pw), respectively, induces in the medium a nonlinear optical polarization (see Eqs. (26)-(27) in Ref.[10]) which is:

$$P_i^{NL} = \chi_{ijk}^{(2)} E_{pj}(\omega_p) E_{pwk}^*(\omega_s) \exp(i[(\omega_p - \omega_{pw})t - (\vec{k}_p - \vec{k}_{pw}) \cdot \vec{r}]) + c.c. \quad (10)$$

where $\chi_{ijk}^{(2)}$ is the susceptibility of rank two tensor components of the crystal. Consequently, such polarization, acting as a source in the wave equation will radiate a new wave $E_i(\omega_i)$ at frequency: $\omega_i = \omega_p - \omega_{pw}$ with an amplitude proportional to $E_{pw}^*(\omega_i)$, i.e., to the complex conjugate of the spatial amplitude of the low-frequency probe wave at $\omega_{pw}$. Moreover, it was shown [12] that a necessary condition for a phase-coherent cumulative buildup of conjugate-field radiation at $\omega_c = \omega_p - \omega_{pw}$ is that the wave vector $\vec{k}_i$ at this new frequency must be equal to $\vec{k}_c = \vec{k}_p - \vec{k}_{pw}$, i.e., we have

$$\vec{k}_c(\omega_p - \omega_{pw}) = \vec{k}_p(\omega_p) - \vec{k}_{pw}(\omega_{pw}) \quad (11)$$

This condition can be satisfied along only one direction in crystal by using the optical anisotropy to compensate for the (linear) refractive index dispersion. In the degenerate case, i.e. $\omega_p = 2\omega_{pw}$ under slowly varying envelope approximation (SVEA) and undepleted pump approximation the wave equation with the nonlinear polarization as source is reduced to the following coupled linear equations:

$$\frac{dE_{pw}^*}{dz} = igE_c \exp[i\Delta kz], \quad \frac{dE_c}{dz} = -igE_{pw}^* \exp[-i\Delta kz], \quad (12)$$

where $g = \omega_{pw}(\mu/\varepsilon)^{1/2} \chi E_p$, and $\Delta \vec{k} = \vec{k}_p - \vec{k}_{pw} - \vec{k}_c$. Solving Eq. (12), with the usual boundary conditions (see Eq. (42) in Jagannath et al [10]), we obtain the following amplification factor:

$$AF = -2i\left(\frac{g^*}{b}\right)\sinh\left(\frac{bL}{2}\right)\exp\left\{-i\frac{\Delta kL}{2}\right\}, \quad b = \{4|g|^2 - (\Delta k)^2\} \quad (13)$$

of the conjugated wave $E_c$ relative to the probe wave $E_{pw}$ at the end of crystal, where L is the crystal thickness. Consequently, the parametric amplification of the phase conjugated field is possible if $|g| \gg \frac{\Delta k}{2}$. So, the relation (13) demonstrate that the three-wave mixing process is capable of yielding an amplified phase-conjugate replica of the input probe wave. Hence, the optical phase conjugation by three-wave mixing (OPC-TWM) [10], help us to obtain a complete proof of the existence of the crossing reactions (6)-(7) as real processes which take place in SPDC crystals when the phase matching conditions (3) are fulfilled. In fact, a rigorous phenomenological crossing symmetric theory for the non degenerate TWM in nonlinear media, which include all three types of second-order susceptibility tensors with their Kleinsman's permutation symmetry, can be developed in similar way with that discussed in Refs. [14,15].

Therefore, as a corollary of the OPC-TWM of the SPDC crystals, we obtained the following important result:

(R.6) *Quantum Mirrors* (QM): *The SPDC crystals acts as real mirrors (quantum mirror) since by OPC-TWM [or equivalently by the crossing processes (6) (or (7))] any signal photon $s(\omega_s, -\vec{k}_s)$ (or idler photon $i(\omega_i, -\vec{k}_i)$) is transformed in an idler photon $i_s(\omega_i, \vec{k}_i)$ (or signal photon $s_i(\omega_s, \vec{k}_s)$), respectively. The high quality of the quantum mirrors is given by the distortion-undoing and amplification properties of these mirrors.*

3.3. *QM interpretations of the entangled photon experiments*

Now, by using the QM-result (R.6), we see that the conceptual "unfolded" version (see Fig. 3 in Ref. [3]) of the "ghost " image experiment can now be understand only on the basis of real physical processes (see (R.1)). Indeed, any signal $s(\omega_s, -\vec{k}_s)$-ray reflected on object will be returned via lens (see again Fig. 3 in Ref. [3]) and BS to BBO crystal where by the crossing symmetric process (6) is "reflected " as an $i_s(\omega_i, \vec{k}_i)$-ray up to the fiber tip detector $D_2$ where the image of the object is observed. Thus, the SPDC-BBO crystal is not only a source of high quality coherent pairs of photons (s) and (i) but also acts as a quantum mirror (quantum phase conjugation mirror (PCM)) for the continuation of a geometric s-ray (or i-ray ) as an induced $i_s$ (or $s_i$)-ray via the crossing processes (6)-(7). Hence, the image of an object will occur as a second-order process in the background produced by other phenomena. Consequently, by coincidence measurements one obtains a separation of the coherent photons can be from this big background. The image is well observed when the aperture, lens, and fiber tip are located according to the Gaussian thin-lens equation (1). In similar way we obtain interpretation of the (R.2) results for the two-photon entangled interference-diffraction patterns [3].

Next, in order to see that our picture of the crystal as a quantum mirror (R.6) (or equivalently as a phase conjugation mirror) is really more powerful, here we apply the QM concept (R.6) to the interpretation of a more recent experiment [16]. In the two-photon imaging experiment [16] the SPDC crystal is illuminated uniformly by a high quality laser pump (p) via the convergent lens (with focal distance f) which now is placed between the pump and SPDC\ crystal at distance d of crystal (see Figs.1 and 6 in Ref. [16]). In few words, our quantum mirror scheme relative to the experiment [16] is as follows: The signal photons from the original SPDC-process (5) are going toward an object P (via a polarizer beam splitter). After an usual reflection on the object P they are returned toward S-QM system (cystal+pump) reaching the pump spherical front of wave in the point A from SPDC-crystal where they are transformed in idler photons by a crossing symmetric reaction (6). In this case, we prove that the system (laser+crystal) behaves like a spherical quantum mirror (S-QM)(see Fig. 2) for which the distance $Z_s$ from the object (P) to the crystal point A and the distance $Z_i$ from the crystal point A to the image point P must satisfy the following important S-QM laws:

(R.7) *S-QM radial laws*

$$\frac{\omega_s}{Z_s} + \frac{\omega_i}{Z_i} = \frac{\omega_p}{R}\cos\beta, \quad \cos\beta = \left[\omega_s \cos\beta_{ps} + \omega_i \cos\beta_{pi}\right]/\omega_p \qquad (14)$$

*with the magnification factor M given by*

$$\mathrm{M} = -\frac{Z_i \sin\beta_{pi}}{Z_s \sin\beta_{ps}} = -\frac{Z_i \omega_s}{Z_s \omega_i} = -\frac{Z_i \lambda_i}{Z_s \lambda_s} \qquad (15)$$

*where $\beta_{ps}$ and $\beta_{pi}$ are the incidence and "reflection" angles (see Fig. 2a,b from Ref. [17]) in the point A upon exiting the crystal.*

The proof of the results (14) and (15) are completely geometrical (see Fig. 2) and they are easy obtained by expliciting the relation: $Area(\Delta PAP') = Area(\Delta PAC) + Area(\Delta AP'C)$ and the conservation laws (3) of the crossing symmetric SPDC process (6), where C is the centre of the sphere with radius CA=R=f-d. Of course, in this proof, the relation $\omega_s \sin \beta_{ps} = \omega_i \sin \beta_{pi}$ (implied by the conservation of the transverse components of the k-vectors in conjuction with Snell's law upon exiting the SPDC-crystal), is taken into account. We note that a detailed proof of the results (14) and (15) as well as the proof for the tangential focal law of the quantum spherical mirror will be presented in Ref. [17].

Therefore, in the particular case of paraxial approximation when $\cos \beta \approx 1$, from the S-QM law (14) we get the following fundamental radial S-QM law

$$\frac{\omega_s}{Z_s(\frac{\lambda_s}{2\lambda_p})} + \frac{\omega_i}{Z_i(\frac{\lambda_i}{2\lambda_p})} = \frac{2}{f-d} = \frac{2}{R} \quad (16)$$

where $\lambda_r, r = p, s, i$ are the corresponding wavelength of the pump, signal and idler photons, respectively.

By inspection we see that Eq. (16) is just the result (20) from Ref. [16] where only this relation (not and magnification factor) was obtained by the minimization of the coincidence counting rate. In our scenario the S=QM radial law (14) and the\ magnification factor (15) are both proved only by using the energy-momentum conservation law (1) of the SPDC-process (5) independent of the fact that are present or not the coincidence circuits. It should be noted that double-coincidence circuits, which can be of kind: (signal, idler), (pump, signal) and (pump, idler), etc., can be of great help for the separation of the coherent photons produced by the crossing symmetric processes (5)-(7) from the background coming from other processes.

Therefore, our results (14) and (15) are not only more complete and more exact but also more general since they can be applied in the following direct S-QM experiment without any kind of coincidence circuit:

(R.8) *A direct QM experiment: In a direct S-QM experiment the object P, in an experimental setup similar to that from Fig.4 of Ref. [17], can be illuminated directly by an independent high quality laser with the same characteristics as that of signal beam from the original SPDC process (5). If the distance $Z_s$ from the object to crystal and the distance $Z_i$ from crystal to image satisfy the law (14), then, the sharp image as well as the magnification factor (15) will be observed in all kind of measurements independent of any kind of coincidence circuits.*

Therefore, by our quantum mirror picture the interpretations of the experiments [3] and [16] are included as a particular case into a more general optical geometric scheme of the kinematical correlated photons in which the proofs of all results (e. g. the results (20) and (22) from Ref. [16]) as well as of the observed magnification factor (15), are obtained independent of the existence of any coincidence circuits. Moreover, our QM-approach is expected to be more powerful and due to the extraordinary properties proved for the optical phase conjugation processes (6)-(7) such as amplification (see the amplification factor in Eq. (13), section 3.2), high coherence, distortion undoing, high resolution, etc. Clearly, all these characteristic features are not present in the two-photon entangled mechanism presented in Refs. [3] and [16].

## Conclusions

The main results and conclusions obtained in this paper can be summarized as follows:
(i) The Cherenkov-like coherence conditions (4) of form $v_{ph}(\omega) \leq v_0$ which are equivalent to

$$\text{Re}\, n_p(\omega_p) \leq \text{Re}\, n_f(\omega_f), \quad f \equiv s, i \quad (17)$$

Therefore, in the signal ( as well as the idler) beams the photons have very high quality coherence properties relative to the pump photons. This coherence condition is similar with that from the usual Cherenkov effect. This important property allow us to determine the refractive properties of crystals when SPDC phenomena can take place;

(ii) We enriched the class of SPDC phenomena by introducing the\ crossing symmetric processes (5)-(6) as real phenomena described just by the same\ transition amplitude as that of the original\ SPDC (5) and satisfying the same energy-momentum conservation law (3). Moreover, the connection between the crossing symmetric reactions (5)-(6) and the well known optical phase conjugation phenomena is established. Such connection is very important not only for a more deeply understanding of the crossing symmetric SPDC processes, but also for a complete explanation of the amplification and control of these phenomena (see Refs.[4-7]);

(iii) On the basis of the complete set (5)-(7) of the SPDC phenomena, as well as on the basis of their high coherence properties (i), in this paper we obtained a new and complete interpretation of the two photon "ghost " image [3], [16] , as well as of the two photon "ghost " interference-diffraction experiments [3]. In few words, the key of this new interpretation is directly connected with the quantum mirror property (see (R.6) in Sect.3) of the SPDC crystals. This new mechanism is also in a very good agreement with the recent experimental results [4-7] on the control of the visibility of these nonlocal second-order phenomena with idler beams or via laser beams aligned with idler beams.

(iv) New and dedicated experiments for a detailed investigations of the crossing symmetric SPDC-phenomena (6)-(7) as well as of the quantum mirror property of the BBO crystals, are needed. For example, as a first class of experiments we suggest that the variation of image (or fringes of the diffraction patterns, etc.) visibility in the idler detector $D_1$ with the power of a signal laser which is illuminating directly (not via SPDC crystal) the object (or the apertures) (see Figs.1 and 3) will prove that the quantum mirror properties (see R.6 and R.7) of the SPDC crystals are responsible for the results of the above experiments [3] and [16]. As a second class of experiments, we suggest that the study of the variation of the images visibility (or diffraction fringes visibility) observed in the idler detector as a function of the power of an auxiliary laser aligned with the signal beam could reveal the fact that these images (or diffraction fringes) are generated by the reflected s photons which are transformed in i photons via the SPDC crossing processes (6). Moreover, since in agreement with (13) we predict that the SPDC crystals can works as amplification setup for the conjugated signal (or idler) photons if the parameter $|k|L > \frac{\pi}{4}$, then the experiments, dedicated to study the variation of the images (or diffraction fringes) visibilities in the above coincidence detection experiments as a function of the crystal thickness L, can be considered as essential tests for our approach.

(v) The formalism developed here can also be directly extended to other Cherenkov-like effects not only in the dielectric nonlinear media [8] but also in the nuclear and hadronic [18-19 ] nonlinear media, since in all such cases the energy-momentum conservation (3), the Cherenkov-like coherence conditions of the form (4), as well as the crossing symmetry of the corresponding $S_{fi}$-matrix, are all fulfilled.

(vi) As direct application of these new results we suggest the quantum photography of objects, quantum holography, etc.

Hence, the main advantages of our approach are : its conceptual simplicity, the theoretical generality as well as, the completeness.

Finally, we hope that our results are encouraging for further theoretical and experimental investigations since a clarification of the role played by optical phase conjugation in the above two photon ghosts imaging (or ghost diffraction) experiments will be of great interest for a real progress in the discovery of the true non local effects in the particle entangled experiments.

A very short version of the "quantum mirror" interpretation of the two-photon ghost imaging and ghost interference-diffraction experiments [2,3] was also published in NIPNE-Scientific Report Bucharest, 1996.

(This paper was published in Romanian Journal of Physics, Vol. **45**, Nos. 1-2,  P. 3-14, Bucharest 2000)